\documentclass[aps,twocolumn,nofootinbib,floatfix,superscriptaddress]{revtex4-2}

\usepackage[english]{babel}
\usepackage[utf8]{inputenc}
\usepackage[T1]{fontenc} 
\usepackage[normalem]{ulem}
\usepackage{amsmath,amssymb,amsfonts,physics,bm,graphicx,xcolor,mathtools,dsfont}
\usepackage[colorlinks=true,linkcolor={red!60!black},citecolor={blue!70!black},urlcolor={blue!60!black}]{hyperref}
\usepackage[nameinlink]{cleveref}
\usepackage{mathrsfs}
\usepackage{xcolor}

\let\oldbibitem\bibitem 
\renewcommand{\bibitem}{
    \renewcommand{\doi}[1]{\texttt{\href{https://doi.org/##1}{doi:##1}}} 
    \let\bibitem\oldbibitem 
    \oldbibitem 
}

\begin{document}

\title{Anisotropic signatures in the spin--boson model}

\author{F. Hartmann}
\email{hartmann3@uni-potsdam.de}
\affiliation{University of Potsdam, Institute of Physics and Astronomy, Karl--Liebknecht--Str. 24--25, 14476 Potsdam, Germany}
\author{S. Scali}
\affiliation{Department of Physics and Astronomy, University of Exeter, Stocker Road, Exeter EX4 4QL, UK}
\author{J. Anders}
\affiliation{University of Potsdam, Institute of Physics and Astronomy, Karl--Liebknecht--Str. 24--25, 14476 Potsdam, Germany}
\affiliation{Department of Physics and Astronomy, University of Exeter, Stocker Road, Exeter EX4 4QL, UK}

\begin{abstract}
Thermal equilibrium properties of nanoscale systems deviate from standard macroscopic predictions due to a non--negligible coupling to the environment. For anisotropic three--dimensional materials, we derive the mean force corrections to the equilibrium state of a classical spin vector. The result is valid at arbitrary coupling strength. 
Specifically, we consider cubic, orthorhombic, and monoclinic symmetries, and compare their spin expectation values as a function of temperature. We underpin the correctness of the mean force state by evidencing its match with the steady state of the simulated non--Markovian spin dynamics. The results show an explicit dependence on the symmetry of the confining material. In addition, some coupling symmetries show a spin alignment transition at zero temperature.
Finally, we quantify the work extraction potential of the mean force--generated inhomogeneities in the energy shells. Such inhomogeneities constitute a classical equivalent to quantum coherences.
\end{abstract}

\maketitle

Standard thermodynamics assumes that the interaction between the system and the bath is negligible compared to the bare system's energy.
Over the last years, much effort has been made to obtain a consistent thermodynamic theory of strongly coupled systems in the classical and quantum regime~\cite{jarzynski2004a,seifert2016,jarzynski2017a,strasberg2017,talkner2020,miller2017,aurell2018,trushechkin2021,thingna2012,campisi2009}.
Here, the thermal equilibrium state is described by the mean force (Gibbs) state, which has been studied comprehensively in the classical and quantum regime for one--dimensional and isotropic three--dimensional interactions~\cite{trushechkin2022,binder2018,chiu2022,cresser2021,anto-sztrikacs2022a}.
It has been shown recently, that for a one--dimensional $\theta$--angled spin--boson model, the quantum mean force state becomes precisely the classical mean force state in the large--spin limit~\cite{cerisola2022}.
This establishes the correspondence principle for an open system for the first time.
Further, environment--induced coherences, so called energy--shell inhomogeneities, are found to be present in the classical mean force state~\cite{smith2022,cerisola2022}.

Meanwhile, magnetic materials with anisotropic crystal structures have been studied in condensed matter physics and magnetism, such as the orthorhombic rare--earth compound DyMnO3~\cite{pekala2013,harikrishnan2009,bhoi2019,bogdanov2007a,antropov2021,biswas2022}.
Further examples are the Mn--doped monoclinic ZrO2 compound~\cite{srivastava2020} or monoclinic Fe3Se4 nanostructures~\cite{long2011,zhang2011,singh2020}.
Effects of the anisotropic crystal geometry lead to differences in the magnetization behaviour with respect to temperature.

In this paper, we consider a three--dimensional classical spin--boson model, where the bath can be anisotropic. 
We give an analytical expression of the classical mean force (CMF) state and study the influence of cubic, orthorhombic and monoclinic crystalline symmetries.
We show that the CMF state is strongly dependent on the symmetry of the coupling.
For the cubic, isotropic bath the CMF state reduces to the classical Gibbs (CG) state.
Further, in the case of orthorhombic crystal symmetry, we find a spin alignment transition at zero temperature.
This results from a change in the potential minimum in the transition from weak to strong system--bath coupling.
The observed classical spin alignment transition shows similarities with the quantum phase transition in the one--dimensional quantum spin--boson model~\cite{anders2007a,alvermann2009,florens2010,defilippis2020a,wang2019b,leggett1987}.

Lastly, anisotropic crystal symmetries lead to classical energy--shell inhomogeneities in the phase space density. 
These were recently linked to a work extraction potential by Smith et al.~\cite{smith2022}.
We show that not all anisotropic baths lead to energy--shell inhomogeneities even though they lead to mean force corrections. 
For the orthorhombic mean force state, we demonstrate that the maximal work extraction is a function of the bath temperature and coupling strength.

In order to answer the question of how different bath symmetries influence the equilibrium properties of a single spin, the paper is organised as follows.
In Section~\ref{sec:setup}, we employ the spin--boson model to calculate the three--dimensional CMF state and use the Neumann principle to construct coupling tensors that represent different crystal symmetries. 
In Section~\ref{sec:results}, we detail the mean force corrections caused by different crystal symmetries.
For orthorhombic crystal symmetries, we observe a spin alignment transition that we discuss in more detail in Section~\ref{sec:spinflip}.
Classical mean force states lead to energy--shell inhomogeneities, that we link to a work extraction potential in Section~\ref{sec:inhomo}. 
We conclude with brief summary and discussion in Section~\ref{sec:discussion}.

\section{Setup}
\label{sec:setup}
The spin--boson model is used in many different physical, chemical and biological contexts~\cite{weiss1999,ferialdi2017,orman2020,lloyd2011,kolli2012,huelga2013,arndt2009}.
For example, it is adopted to describe the dissipation and decoherence effects in graphene~\cite{dattagupta2021} and to study the heat transfer in non--equilibrium situations~\cite{wang2017a}. 
Further, environment induced quantum phase transitions between delocalized and localized states are observed in the spin--boson model~\cite{defilippis2020,zhang2010} and it describes the physics of quantum emitters that are coupled to surface plasmons~\cite{dzsotjan2010}.

\subsection{3D spin--boson model}
\begin{figure}
    \centering
    \includegraphics[width=0.36\textwidth]{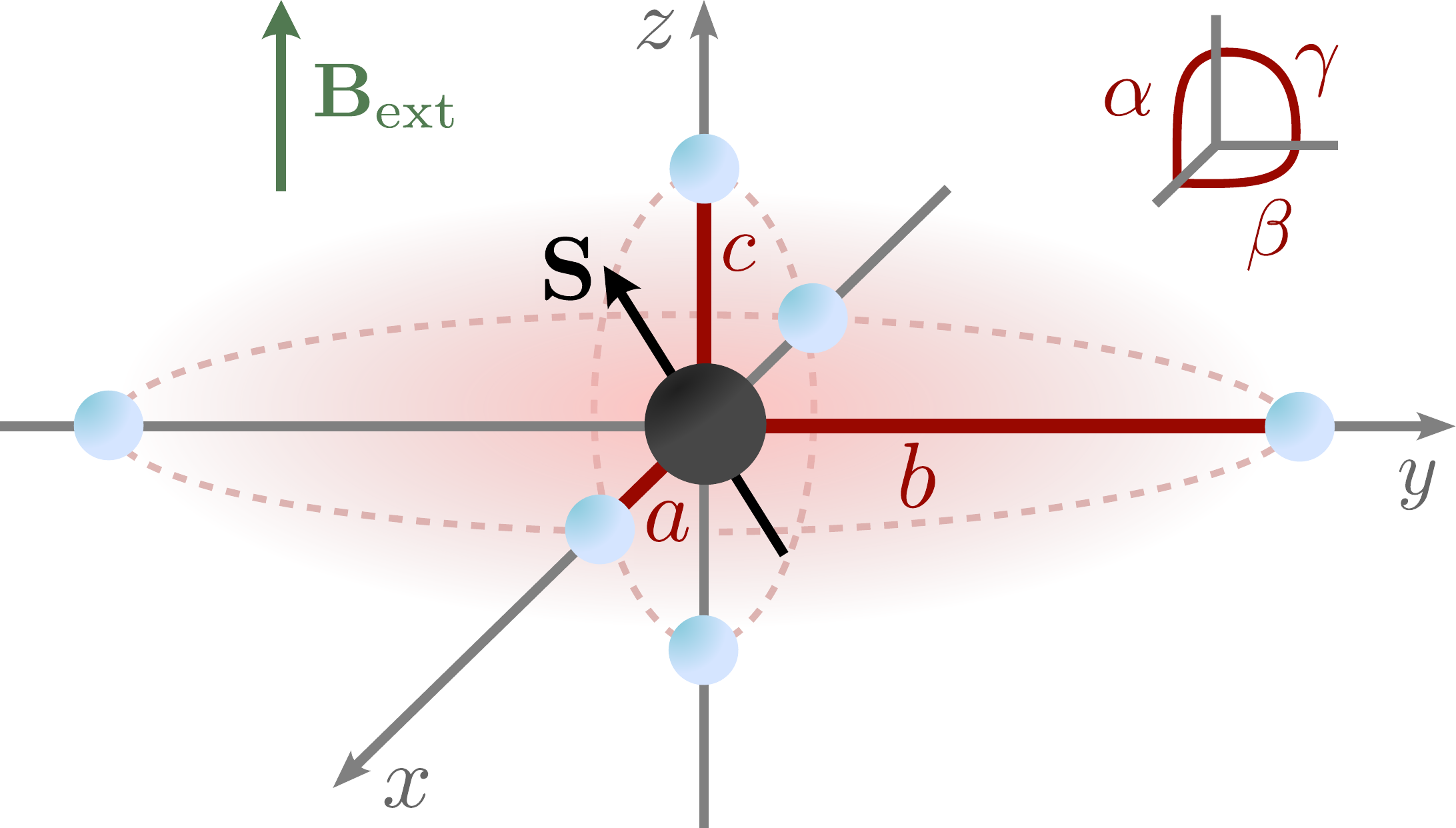}
    \caption{\textbf{Sketch of a lattice giving rise to an anisotropic noise field:} The spin (indicated in black) is surrounded by a bath of neighbouring atoms (shown in light blue) forming an anisotropic lattice. Lattice vibrations (phonon excitations) at finite temperatures lead to an anisotropic noise field (red shade). Differences in lattice geometries result in different spin--bath interactions and thus to corrections to the spin's equilibrium state. The lattice parameters $a$, $b$, and $c$ together with the angles $\alpha$, $\beta$, and $\gamma$ characterize different crystal classes, captured in the tensor $\mathcal{C}$. We assume the external magnetic field $\mathbf{B}_\mathrm{ext}$ (green vector) to be aligned along the $z$--direction.}
    \label{fig:SketchModel}
\end{figure}

Here, we introduce the system which is composed of a classical single spin vector $\mathbf{S}$ with length $S_0$, exposed to an external magnetic field $\mathbf{B}_{\mathrm{ext}} = (0,0,B_z)$, with $\omega_{\mathrm{L}} = |\gamma| B_z$ being the Larmor frequency (see sketch in Fig.~\ref{fig:SketchModel}).
The system Hamiltonian is given by
\begin{equation}
H_{\mathrm{S}} = -|\gamma|\mathbf{S}\cdot\mathbf{B}_{\mathrm{ext}} = -\omega_{\mathrm{L}}S_z.
\end{equation}
The spin is embedded into a thermal bosonic bath. 
Even though we assume the bosonic bath to consist of phonons in a crystal lattice~\cite{nemati2022}, one could also consider modes of an electromagnetic field. 
We model the phonon modes by the Hamiltonian
\begin{equation}
H_{\mathrm{B}} = \frac{1}{2}\int_0^{\infty}\mathrm{d}\omega\,\big(\mathbf{P}_\omega^2 + \omega^2\mathbf{X}_\omega^2\big),
\end{equation}
where $\mathbf{P}_\omega$ and $\mathbf{X}_\omega$ are the three--dimensional phase space coordinates of a mode with frequency $\omega$.
The interaction between the system and the bath is assumed to be linear, which is sensible in most settings~\cite{trushechkin2022}.
Hence, the interaction Hamiltonian is modelled as
\begin{equation}
H_{\mathrm{int}} = \mathbf{S}\cdot\int_0^\infty\mathrm{d}\omega\,\mathcal{C}_\omega\mathbf{X}_\omega,
\end{equation}
with $\mathcal{C}_\omega$ being the coupling tensor that determines the strength and the symmetry properties of the system--bath interaction.
Unlike previous investigations establishing the link to the LLG equation~\cite{anders2022} and the first open system quantum--classical correspondence~\cite{cerisola2022}, in this work, we investigate the physical implications of the $(3\times3)$ second rank coupling tensor $\mathcal{C}_\omega$~\cite{nemati2022}.

The total Hamiltonian of the combined system--bath compound is given by~\cite{anders2022,cerisola2022},
\begin{equation}
H_{\mathrm{tot}} = H_{\mathrm{S}} + H_{\mathrm{B}} + H_{\mathrm{int}}.
\end{equation}
This Hamiltonian guides the dynamical evolution and the equilibrium features of the spin system interacting with a classical harmonic environment.
The mean force state is generally defined as~\cite{trushechkin2022},
\begin{equation}
\tau_{\mathrm{MF}} \coloneqq \mathrm{tr}^{\mathrm{cl}}_{\mathrm{B}}[\tau_{\mathrm{tot}}] = \mathrm{tr}_{\mathrm{B}}^{\mathrm{cl}} \bigg[\frac{e^{-\beta H_{\mathrm{tot}}}}{Z_{\mathrm{tot}}}\bigg].
\end{equation}
This is the reduced state of the global Gibbs state $\tau_{\mathrm{tot}}$, with the global partition function $Z_{\mathrm{tot}}$, the inverse temperature $\beta = 1/k_{\mathrm{B}}T$, where $k_{\mathrm{B}}$ is the Boltzmann constant. 
Taking the partial trace in the classical setting, $\mathrm{tr}_{\mathrm{B}}^{\mathrm{cl}}[\cdot]$, requires integrating over the bath degrees of freedom $(\mathbf{X}_\omega,\mathbf{P}_\omega)$~\cite{cerisola2022}.
In the Appendix~\ref{app:MFder}, we show a detailed derivation and give the exact definitions of the classical partial trace of the bath and the spin system.

\subsection{3D classical mean force state (CMF)}
Carrying out the trace over the bath, we derive the three-dimensional CMF state for arbitrary coupling strengths:
\begin{equation}
\tau_{\mathrm{MF}} = \frac{1}{\tilde{Z}_{\mathrm{S}}^{\mathrm{cl}}}e^{-\beta\big(H_\mathrm{S}-\frac{1}{2}\int_0^\infty\mathrm{d}\omega(\mathbf{S}^{\mathrm{T}}\mathcal{C}_\omega\mathcal{C}_\omega^{\mathrm{T}}\mathbf{S})/\omega^2\big)},
\label{eq:meanforce0}
\end{equation}
where $\tilde{Z}_{\mathrm{S}}^{\mathrm{cl}} = \mathrm{tr}_{\mathrm{S}}^{\mathrm{cl}}[\exp{-\beta(H_\mathrm{S}-\frac{1}{2}\int_0^\infty\mathrm{d}\omega(\mathbf{S}^{\mathrm{T}}\mathcal{C}_\omega\mathcal{C}_\omega^{\mathrm{T}}\mathbf{S})/\omega^2)}]$ is the spin's CMF partition function.
This is the first result of the paper and is an upgrade of the one--dimensional CMF state discussed in~\cite{cerisola2022}. 
Our result is valid for any three--dimensional bath symmetry and all coupling strengths.
It remains an open task to find a closed expression for the quantum mean force state for all coupling symmetries and coupling strengths~\cite{trushechkin2022,cresser2021,anto-sztrikacs2022a}.

From here onwards we will assume that the frequency dependence given by the spectral density $J_\omega$ is isotropic,
\begin{equation}
\mathcal{C}_\omega = \sqrt{2\omega J_\omega}\cdot\mathcal{C}.
\label{eq:Spectral}
\end{equation}
But the overall strength of coupling can vary in different spatial directions, which is set by the elements of the tensor $\mathcal{C}$ (Sec.~\ref{subsec:crystalsymmetry}).
In what follows, we assume that the spectral density of the lattice vibrations is of Lorentzian form $J_\omega = (A\Gamma\omega)/(\pi(\omega_0^2-\omega^2)^2 + \pi\Gamma^2\omega^2)$. 
This is a reasonable assumption thanks to the bosonic spectral density $J_\omega$ being proportional to the phononic density of states $D_\omega$, i.e. $J_\omega\omega \propto D_\omega$. Note that Lorentzian shaped spectral densities are a good choice to describe experimentally measured $D_\omega$~\cite{nemati2022}.
Further, a Lorentzian spectral density has the advantage to lead to a closed set of differential equations when simulating the spin dynamics~\cite{cerisola2022,anders2022}.
Under these assumptions, the frequency integral in the exponent of Eq.~\eqref{eq:meanforce0} then simplifies to the reorganization energy $Q = A/(2\omega_0^2)$.
While the dynamics is governed by the peak width $\Gamma$, the eventual equilibrium state only depends on the resonant frequency $\omega_0$ and the spin--bath coupling strength $A$~\cite{anders2022}.
Further, we rename the spin--matrix product $\tilde{S}^2 =\mathbf{S}^{\mathrm{T}}\mathcal{C}\mathcal{C}^{\mathrm{T}}\mathbf{S}$ such that the CMF state simplifies to~\cite{cerisola2022},
\begin{align}
\tau_{\mathrm{MF}}(\vartheta,\varphi) = \frac{1}{\tilde{Z}_{\mathrm{S}}^{\mathrm{cl}}} e^{-\beta(H_{\mathrm{S}} - Q\tilde{S}^2)}. 
\label{eq:meanforce}
\end{align}
In Eq.~\eqref{eq:meanforce}, $\tau_{\mathrm{MF}}(\vartheta,\varphi)$ is a distribution given in terms of the angles of the spherical coordinates $(\vartheta,\varphi)$, where $\mathbf{S} = S_0(\sin\vartheta\cos\varphi,\sin\vartheta\sin\varphi,\cos\vartheta)$ for a fixed spin length $S_0$.

The CMF state is postulated to be the thermal equilibrium state of a spin in contact with a bath, i.e. the dynamical classical steady state (CSS) in the long--time limit. 
While the correspondence of the CMF state and the CSS is proven in the weak coupling limit~\cite{mori2008} and the ultra--strong coupling regime~\cite{trushechkin2021} of quantum systems, there remain some open questions about the formal proof that this is valid for all coupling strengths and coupling symmetries~\cite{trushechkin2022,anto-sztrikacs2022a}.
Here, we demonstrate that the dynamics of a classical spin converges to the CMF state for classical noise, arbitrary coupling strength, and especially, any crystalline coupling symmetry, see Fig.\ref{fig:Monoclinic13Results}(a).
We do this by numerically solving the spin dynamics equations with the analytical and numerical methods detailed in Refs.~\cite{anders2022,scali}.

\subsection{Crystal Symmetries} 
\label{subsec:crystalsymmetry}
Coupling the spin to a harmonic bath that reflects the symmetry of the confining material requires knowledge about the coupling tensors $\mathcal{C}$. 
The specific form of the coupling tensor can be deduced from the Neumann principle~\cite{neumann1885,haussuh2007}. 
The Neumann principle arises from symmetry considerations and connects the structure of a given crystal with its physical properties~\cite{neumann1885,malgrange2014,newnham2005}.
As a result, the coupling tensor must exhibit the same symmetry as the crystal it describes.
This leads to the intuitive observation that, for crystals with more symmetries, the number of independent tensor components decreases. 
In general, the Neumann principle only determines the minimum number of symmetries of the coupling tensor.
The specific form of the tensors that arise from the Neumann principle are discussed in the following by their contribution to the CMF state. 

So far, we solely discussed the coupling tensor $\mathcal{C}$, however, we observe that, in Eq.~\eqref{eq:meanforce}, the symmetric product $\mathcal{C}\mathcal{C}^{\mathrm{T}}$ is responsible for the mean force corrections.
We restrict the product of the matrices $\mathcal{C}\mathcal{C}^{\mathrm{T}}$ to follow the Neumann principle.
In what follows, we refer to the components of $\mathcal{C}\mathcal{C}^{\mathrm{T}}$, as
\begin{equation}
    \mathcal{C}\mathcal{C}^{\mathrm{T}} = \left( \begin{matrix} c_{11}&c_{12}&c_{13}\\c_{12}&c_{22}&c_{23}\\c_{13}&c_{23}&c_{33}\end{matrix} \right).
    \label{eq:proptensor}
\end{equation}
Here, $\mathcal{C}\mathcal{C}^{\mathrm{T}}$ is a symmetric, positive--definite tensor. \\
Different crystal classes are additionally characterized via their lattice parameters $a,\,b,\,c$, and the angles $\alpha,\,\beta,\,\gamma$, as indicated in Fig.~\ref{fig:SketchModel}.

\section{Anisotropic mean force corrections}
\label{sec:results}
In this section, we summarize the effects that a three--dimensional bath with a given lattice structure has on the CMF state of the spin.
In detail, we discuss cubic, orthorhombic and monoclinic crystal symmetries. 
For comparability of the different crystal symmetries we always set the trace of the diagonalized property tensor to unity, i.e. $\mathrm{tr}[\mathcal{C}\mathcal{C}^{\mathrm{T}}] = 1$.

\subsection{Cubic}
\label{subsec:cubic}
A cubic crystal symmetry, i.e. $\alpha = \beta = \gamma = 90^{\circ}$ and $a = b = c$, results via the Neumann principle in an isotropic harmonic bath, i.e. the property tensor simplifies to $\mathcal{C}\mathcal{C}^{\mathrm{T}} = (1/3)\mathds{1}_3$. 
It follows directly from the isotropy of the bath that the classical mean force state reduces to the classical Gibbs (CG) state, $\tau_{\mathrm{MF}}^{\mathrm{cubic}} = \tau_{\mathrm{Gibbs}} = e^{-\beta H_\mathrm{S}}/\tr^\mathrm{cl}[e^{-\beta H_\mathrm{S}}]$, since $\tilde{S}^2 = (1/3)S_0^2$ is constant and independent of $(\vartheta,\varphi)$.
Thus, we observe that classical isotropic three--dimensional noise leaves the Gibbs state invariant with respect to any system--bath coupling strength $Q$. 
\begin{figure*}
	\centering
	\includegraphics[width=\textwidth]{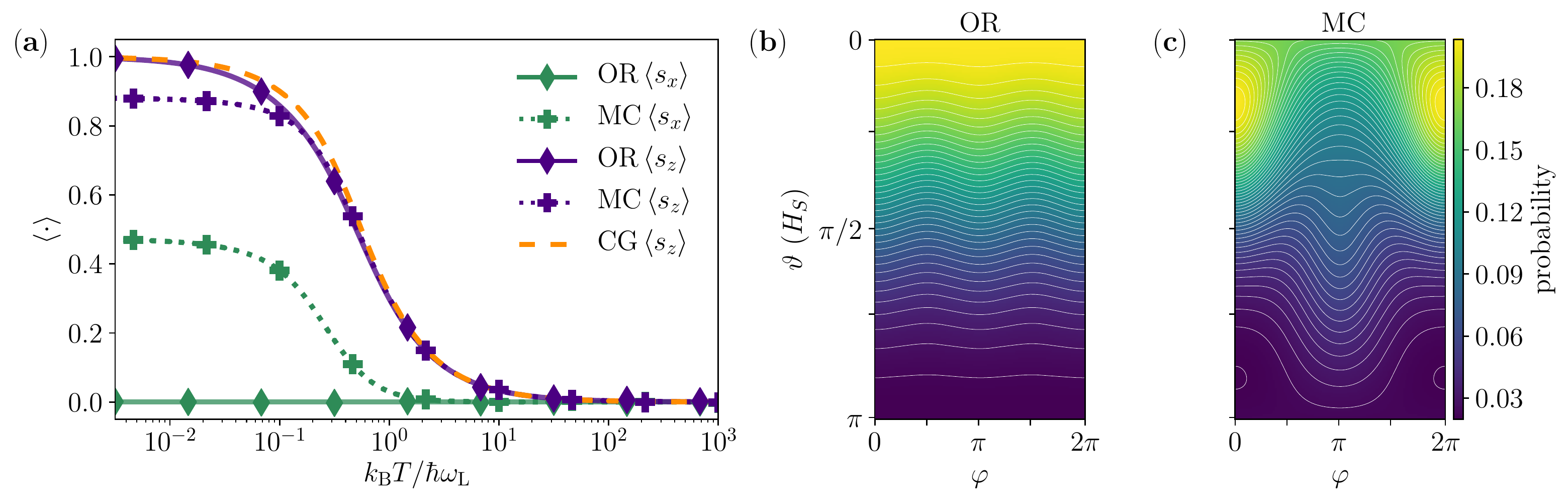}
	\caption{\textbf{CMF results for the orthorhombic and the monoclinic symmetry:} Expectation values of the orthorhombic CMF state (OR) and the monoclinic CMF state (MC) (panel (a)). 
    We observe deviation from the CG state in $\langle s_z \rangle$ for both symmetries (purple solid and dotted line) and a non--zero expectation value $\langle s_x \rangle$ for the monoclinic CMF state (green dotted line). The results are supported by the corresponding dynamical CSS calculations (markers).
    The cubic CMF state is identical to the CG state (orange dashed line). 
    In panel (b), we show the phase space distribution of the orthorhombic CMF state, and in panel (c) of the monoclinic CMF state. 
    The white lines indicate shells of equal probability. One sees that $\tau_{\mathrm{MF}}$ is an inhomogeneous distribution over the energy--shells of $H_{\mathrm{S}}$ for the orthorhombic and monoclinic symmetry, i.e. the white lines are not constant along $\varphi$. 
    The expectation values and distributions are plotted for a temperature of $k_{\mathrm{B}}T = \hbar\omega_{\mathrm{L}}$, a reorganization energy of $Q = 2.5\,\omega_{\mathrm{L}}\hbar^{-1}$, and a spin length of $S_0 = \hbar$.}
	\label{fig:Monoclinic13Results}
\end{figure*}

\subsection{Orthorhombic (OR)}
\label{subsec:ortho}
Orthorhombic crystal symmetries have the following features $a \neq b \neq c$ and $\alpha = \beta = \gamma = 90^{\circ}$. 
This breaks the isotropy of the cubic crystal in the sense that neighbouring atoms have different separations along orthogonal spatial directions. 
Hence, the property tensor of the coupling function has only diagonal elements.
As an example, in Fig.~\ref{fig:Monoclinic13Results}, we choose $c_{11} = 0.40,\,c_{22} = 0.35,\, c_{33} = 0.25$ and the reorganization energy in the strong coupling regime with $Q = 2.5\,\omega_{\mathrm{L}}\hbar^{-1}$~\cite{cerisola2022}. 

The spin expectation values in the $x$ and $y$ directions are $\langle s_x \rangle  = \langle s_y\rangle = 0$ and correspond to the CG state.
This arises because the applied noise is invariant under rotation by $\varphi  = \pi$ around the $z$--axis. 
On the other hand, while the expectation value $\langle s_z \rangle$ for the CMF state slightly deviates from the one corresponding to the CG state,
for $T \to 0\,\mathrm{K}$, both reach $\langle s_z \rangle = 1.0$, see Fig.~\ref{fig:Monoclinic13Results}(a).

For higher coupling strengths, e.g. $Q = 10\,\omega_{\mathrm{L}}\hbar^{-1}$, we observe a bump in the expectation value, as we show in the Appendix~\ref{subsec:bumpyexp}.
In many--body systems, such bumpy magnetization vs. temperature measurements are encountered in materials that show a phase transition to anti--ferromagnetic ordering~\cite{pekala2013,harikrishnan2009,bhoi2019,bogdanov2007a,biswas2022}.
In App.~\ref{subsec:bumpyexp}, we plot the temperature dependent magnetization experiment of an orthorhombic DyMnO3 single crystal from Ref.~\cite{pekala2013} together with the bumpy expectation value of the orthorhombic CMF state.

In Fig.~\ref{fig:Monoclinic13Results}(b), we show the CMF phase space distribution in the case of the OR symmetry.
The contour white lines indicate shells of constant probability.
Deviations from the CG state are obvious, since $\tau_{\mathrm{Gibbs}}$ would have straight parametric lines, i.e. no $\varphi$ dependence.
In contrast, for $\tau_{\mathrm{MF}}^{\mathrm{OR}}$ a teeth--like pattern is formed. 
It arises from the fact that we set $c_{11} > c_{22}$, and therefore, the spin components along the positive and negative $x$--direction $(\varphi = 0,\,\pi)$ are weighted stronger than the components along the $y$--direction.

\subsection{Monoclinic (MC)}
\label{subsec:mono}
Many features of classical mean force corrections are already observed by considering the orthorhombic crystal symmetry.
However, by coupling to a monoclinic bath, further properties are discovered.
In general, a monoclinic crystal structure is characterised via $a \neq b \neq c$ and $\alpha = \beta = 90^{\circ} \neq \gamma$. 
This leads to a coupling tensor in Eq.~\eqref{eq:proptensor} where we set the off--diagonal element $c_{13} \neq 0$ and fix the diagonal elements as for the orthorhombic symmetry.
This geometrically corresponds to rotating the neighbouring atoms in the $\pm x$--direction towards the $z$--axis.
It leads to a correction in the CMF state, i.e. $\tilde{S}^2 = c_{11}S_x^2 + c_{22}S_y^2 + c_{33}S_z^2 + 2c_{13}S_xS_z$, which has a cross--term of the spin--components $S_xS_z \propto \cos\varphi\sin\vartheta\cos\vartheta$. 
This leads to a broken rotational invariance.
We observe a CMF phase space distribution (Fig.~\ref{fig:Monoclinic13Results}(c)) that has a maximum rotated away from the positive $z$--direction, towards the positive $x$--direction. Further, along the negative $x$--direction there is an increased probability to find the spin with a negative $S_z$--component. 
This also leads to a non--zero expectation value $\langle s_x \rangle \neq 0$ for small temperatures $k_{\mathrm{B}}T/\hbar\omega_{\mathrm{L}} < 1$ (see Fig.~\ref{fig:Monoclinic13Results}(a)).

We conclude that, the equilibrium properties of a classical spin vector drastically change when it is anisotropically coupled to a bath. 
In fact, its equilibrium features are highly dependent on the coupling strength, the temperature, and the crystal symmetry, i.e. the coupling tensor $\mathcal{C}\mathcal{C}^{\mathrm{T}}$.

\section{Spin Alignment Transition}
\label{sec:spinflip}
We want to study the classical spin expectation value $\langle s_z \rangle$ at zero temperature while increasing $Q$.
\begin{figure}
    \centering
    \includegraphics[width=0.47\textwidth]{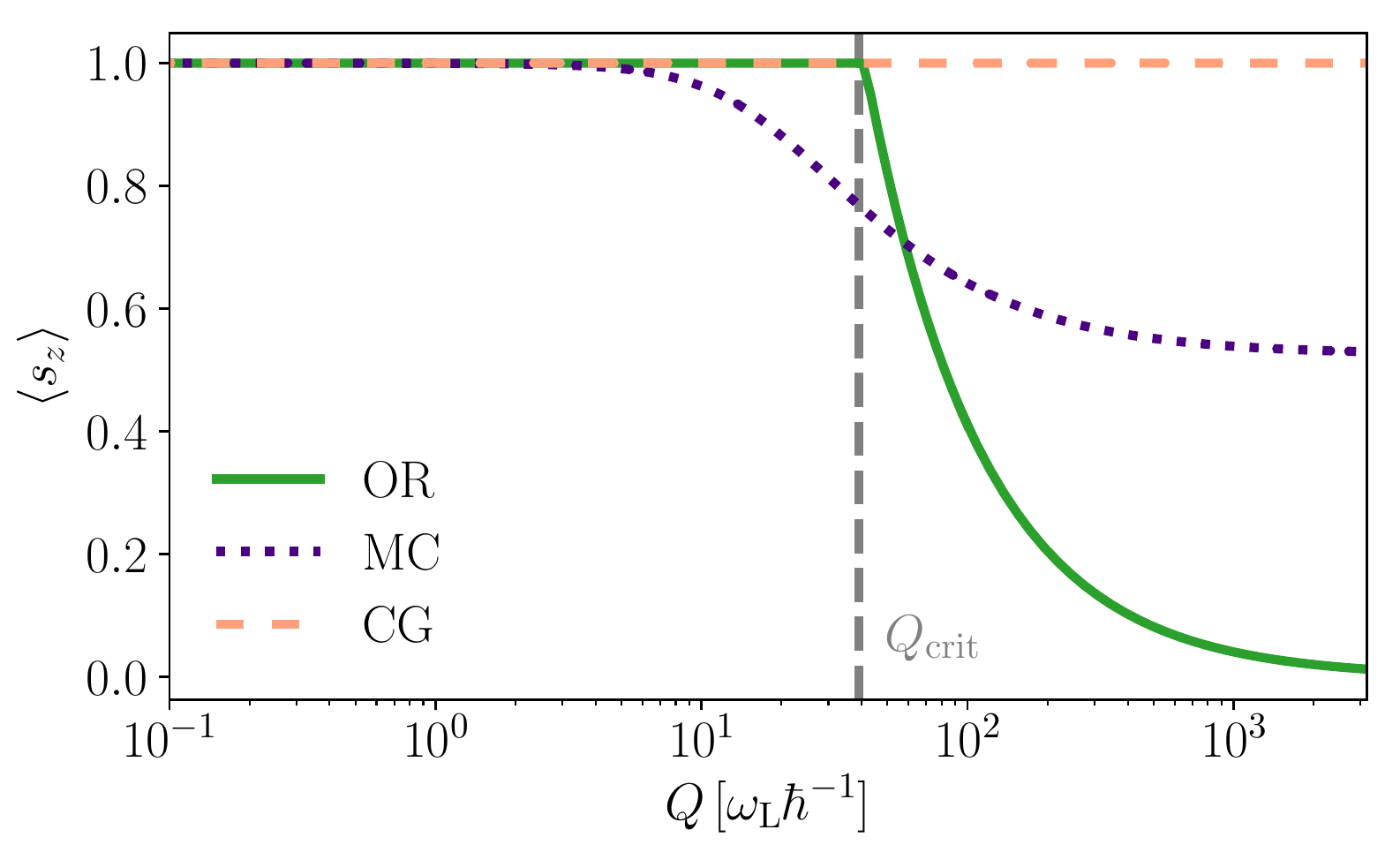}
    \caption{\textbf{Spin alignment transition in classical 3D anisotropic spin--boson model:} $s_z$ expectation value for the orthorhombic (OR, green solid line) and monoclinic (MC, purple dotted line) crystal symmetry at $T \to 0\,\mathrm{K}$ over coupling strength $Q$. 
    The standard Gibbs state (orange) is of course independent of $Q$. 
    The vertical grey--dashed line indicates the critical coupling strength $Q_{\mathrm{crit}}^{\mathrm{OR}}$ at which the transition occurs in the orthorhombic CMF state.
    At $Q_{\mathrm{crit}}^{\mathrm{OR}}$ the most probable alignment of the spin changes from along the $z$--axis towards the $\pm x$--axes.
    In contrast, the monoclinic expectation value reduces continuously to a finite expectation value $\langle s_z \rangle > 0$.
    }
    \label{fig:spinflip}
\end{figure}
Figure~\ref{fig:spinflip} shows the numerically calculated expectation values for the cubic, orthorhombic, and monoclinic symmetry at $T = 0\,\mathrm{K}$.
For the three--dimensional orthorhombic CMF state (OR), we observe a spin--alignment transition, where the expectation value at a critical coupling strength $Q_{\mathrm{crit}}^{\mathrm{OR}}$ abruptly reduces from $\langle s_z \rangle  = 1$ to $\langle s_z \rangle = 0$.
The transition is characterized by minimizing the energy $H^{\mathrm{OR}}_{\mathrm{MF}}(\vartheta,\varphi) = H_{\mathrm{S}} - Q\tilde{S}^2$, where we observe a repositioning of the energetic potential minimum,
\begin{equation}
    \vartheta_{\mathrm{min}}^{\mathrm{OR}} = 
    \begin{cases}
        0 & \quad y < 1 \\
        \arccos\left(\frac{1}{y}\right) & \quad y \geq 1,
    \end{cases}
\end{equation}
with $y = Q/Q_{\mathrm{crit}}^{\mathrm{OR}}$ for $c_{11} > c_{33}$.
The critical coupling strength is given by $Q_{\mathrm{crit}}^{\mathrm{OR}}=\omega_{\mathrm{L}}/(2(c_{11}-c_{33})S_0)$, as indicated in Fig.~\ref{fig:spinflip} by the grey--dashed vertical line.
The azimuthal angle $\varphi_{\mathrm{min}}$ of the energetic potential minimum becomes $\varphi_{\mathrm{min}} = 0, \pi$ for $y \geq 1$.
In other words, the energetic potential minimum becomes degenerate for coupling strengths larger than $Q_{\mathrm{crit}}^{\mathrm{OR}}$. In the zero temperature limit, $\beta \to \infty$, the orthorhombic 
expectation value is 
\begin{equation}
     \langle s_z\rangle_{\beta \to \infty} =
     \begin{cases}
         1 & \quad Q<Q_{\mathrm{crit}}^{\mathrm{OR}} \\
         Q_{\mathrm{crit}}^{\mathrm{OR}}/Q & \quad Q \geq Q_{\mathrm{crit}}^{\mathrm{OR}}.     
     \end{cases}
     \label{eq:analyticdisontinues}
\end{equation}
It is clear that the first derivative of~\eqref{eq:analyticdisontinues} with respect to $Q$ is discontinuous at $Q_{\mathrm{crit}}^{\mathrm{OR}}$, and the expectation value $ \langle s_z\rangle_{\beta \to \infty} $ changes quickly from 1 to 0 for increasing $Q$. At the critical coupling strength it becomes energetically more convenient for the classical spin vector $\mathbf{S}$ to align along the positive and negative $x$--axes, with a consequent zero expectation value along the $z$--direction, $\langle s_z \rangle = 0$.
We note that this spin--alignment transition of a classical spin vector resembles the well--known quantum phase transition in the one--dimensional quantum spin--boson model~\cite{leggett1987,anders2007a,alvermann2009,florens2010,Zhou_2015,defilippis2020a,wang2019b}. 
This raises the question how the quantum phase transition is influenced by three--dimensional anisotropic baths in the quantum spin--boson model.
This could be studied with the recently developed FP--HEOM approach which is robust at zero temperature~\cite{xu2022}.
In contrast to the OR case, the monoclinic CMF state (MC) shows a smooth transition from $\langle s_z \rangle = 1$ to a finite value $\langle s_z \rangle > 0$ for increasing $Q$.

We conclude that the observed classical spin--alignment transition at zero temperature is highly dependent on the anisotropy of the system--bath coupling.

\section{Work extraction potential}
\label{sec:inhomo}
\begin{figure*}
    \centering
    \includegraphics[width=\textwidth]{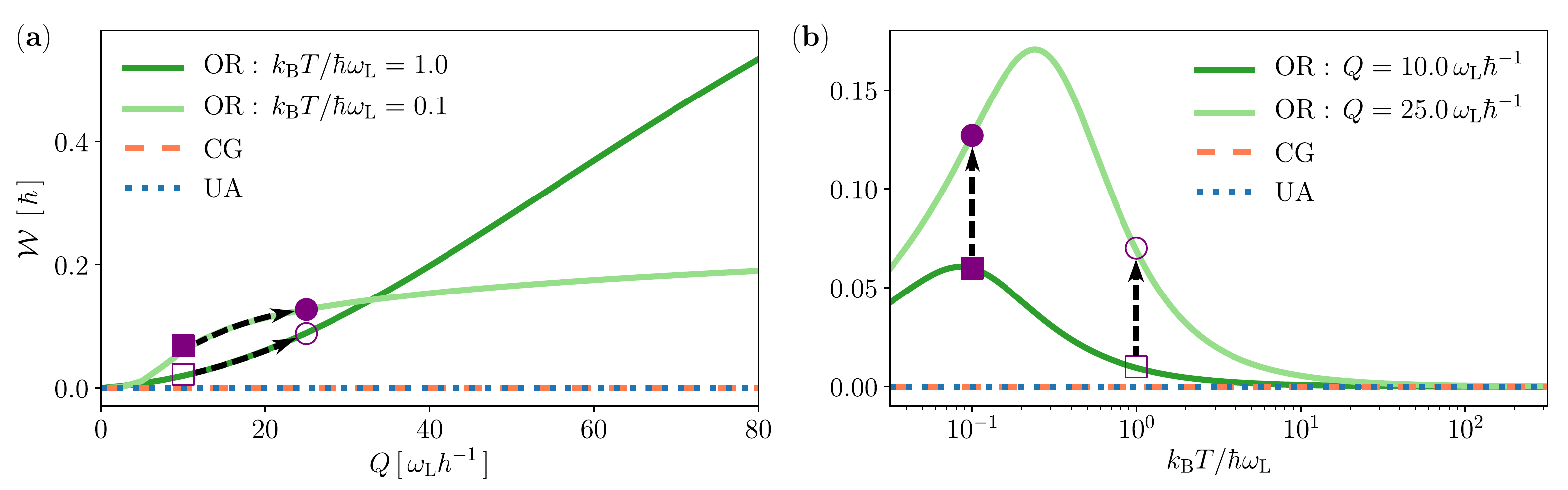}
    \caption{\textbf{Work extraction potential for different crystal symmetries:}
    (a) $\mathcal{W}$ for the orthorhombic crystal symmetry as a function of the reorganization energy $Q$, for two different temperatures (solid green lines).
    For the cubic (orange--dashed) and uniaxial (blue--dotted) symmetry $\mathcal{W}$ is zero, due to missing energy--shell inhomogeneities. 
    (b) Same as (a) but now shown over temperature $T$, for two values of $Q$ (solid green lines).
    At a fixed coupling strength, there exists an optimal temperature, at which the work extraction is maximized (e.g. if $Q = 25\,\omega_{\mathrm{L}}\hbar^{-1}$ the maximal work can be extracted at a temperature of $k_{\mathrm{B}}T/\hbar\omega_{\mathrm{L}} \approx\,0.3$).
    This shows that the trade--off between the temperature and the coupling strength is non--trivial.
    The purple filled (unfilled) squares and circles are plotted for easier comparison between (a) and (b). 
    Mind the semi--logarithmic scaling in the right figure. The classical spin length is $S_0 = \hbar$.}
    \label{fig:work}
\end{figure*}
CMF corrections in the presence of anisotropic system--bath coupling can introduce energy--shell inhomogeneities~\cite{cerisola2022}, see Fig.~\ref{fig:Monoclinic13Results}(b) and (c).
Recently Smith et al.~\cite{smith2022} showed, that classical inhomogeneities are equivalent to quantum coherences as a thermodynamic resource for work extraction~\cite{kammerlander2016}.
Here, we link CMF states with anisotropic coupling to such a work extraction potential.
Similar to the idea of extracting work from quantum coherences by altering the state with coherences to a state without coherences, one can extract work from classical distributions by removing energy--shell inhomogeneities~\cite{smith2022}.
The energy--shells are latitude circles, described by $H_{\mathrm{S}}(\vartheta)$, where the spin phase space is a sphere with radius $S_0$.
The maximal work extraction is given by~\cite{smith2022}, 
\begin{equation}
    \mathcal{W} = \beta^{-1}\big(S[\mathrm{diag}\,\tau_{\mathrm{MF}}(\vartheta)] - S[\tau_{\mathrm{MF}}(\vartheta,\varphi)]\big),
    \label{eq:work}
\end{equation}
where, as before, $\tau_{\mathrm{MF}}(\vartheta,\varphi)$ is the CMF state which may host energy--shell inhomogeneities, and $\mathrm{diag}\,\tau_{\mathrm{MF}}(\vartheta) = \int\mathrm{d}E\,\eta(E)\,\omega_{E}(\vartheta,\varphi)$ defines the homogeneous distribution. Here, $\eta(E)$ is the distribution of energies of a phase space density $\rho(\Gamma)$ and $\omega_{E}(\cdot)$ is the classical microcanonical density.
In addition, $S[\rho]$ refers to the Shannon entropy of a probability distribution $\rho$.

Following~\cite{smith2022}, we calculate the homogeneous distribution of the orthorhombic CMF state,
\begin{align}
    \mathrm{diag}\,\tau_{\mathrm{MF}}^{\mathrm{OR}}(\vartheta) =& \frac{1}{\tilde{Z}_{\mathrm{S}}^{\mathrm{cl}}}\exp\Big[-\beta\big(H_{\mathrm{S}} - QS_0^2F(\vartheta)\big)\Big]\nonumber
    \\
    &\times\mathrm{I}_0\Big[\frac{1}{2}\beta QS_0^2(c_{11} - c_{22}) \sin^2\vartheta\Big],
    \label{eq:orthohomo}
\end{align}
where $F(\vartheta) = \frac{1}{2}\sin^2\vartheta(c_{11} + c_{22}) - c_{33}\cos^2\vartheta$, and $\mathrm{I}_0(x)$ is the modified Bessel function of the first kind.
A detailed derivation is given in Appendix~\ref{sec:appendix_inhomogeneities}.
Eq.~\eqref{eq:orthohomo} is solely dependent on $\vartheta$ and, hence, is homogeneous in the energy--shells $H_{\mathrm{S}}(\vartheta)$.

Following~\cite{smith2022}, in order to calculate the classical work potential $\mathcal{W}$, one needs a coarse--grained phase space distribution.
The probability of the $k^\mathrm{th}$ cell is given by,  $p_k = \int_{k^\mathrm{th}\,\mathrm{cell}}\mathrm{d}\Omega\,\tau_{\mathrm{MF}}(\vartheta,\varphi)$ and $0 \leq p_k \leq 1$.
Hence, the entropy in Eq.~\eqref{eq:work} is calculated via,
\begin{align}
    S[\rho] = -\sum_k p_k\ln{p_k},
\end{align}
for which $S[\rho] \geq 0$.
We want to compare the work potential of the orthorhombic case, by introducing a CMF state $\tau_{\mathrm{MF}}^{\mathrm{UA}}(\vartheta) \propto \exp\left[-\beta(H_{\mathrm{S}} - QS_0^2(c_{11}\sin^2\vartheta + c_{33}\cos^2\vartheta)\right]$, where the coupling to the bath is of uniaxial (UA) symmetry. For this symmetry, $a = b \neq c$ and $\alpha = \beta = \gamma = 90^{\circ}$, and the non--trivial elements in the tensor $\mathcal{C}\mathcal{C}^{\mathrm{T}}$ (Eq.~\eqref{eq:proptensor}) are $c_{11} = c_{22} \neq c_{33}$. 
Even though the UA crystal symmetry introduces mean force corrections, the CMF state $\tau_{\mathrm{MF}}^{\mathrm{UA}}(\vartheta)$ has no energy--shell inhomogeneities, since it exclusively depends on $\vartheta$.
We can also view Eq.~\eqref{eq:work}, as a measure of how inhomogeneous a given CMF state is for different system--bath symmetries, coupling strengths, and temperatures.

In Fig.~\ref{fig:work}, we compare the work extraction potential $\mathcal{W}$ for different crystal symmetries. 
As expected, the isotropic, cubic crystal symmetry leads to $\mathcal{W} = 0$ (orange--dashed line). 
The same is observed for the uniaxial CMF state (blue--dotted line), since its mean force corrections do not host energy--shell inhomogeneities. 
But for the orthorhombic crystal symmetry a non--zero potential arises due to $\mathrm{diag}\,\tau_{\mathrm{MF}}^{\mathrm{OR}}(\vartheta) \neq \tau_{\mathrm{MF}}^{\mathrm{OR}}(\vartheta,\varphi)$.

In Fig.~\ref{fig:work}(a), we observe the following trend. At small $Q$, lower temperatures (solid light green) correspond to greater work extraction potential. Notably, we find the opposite behaviour at increasing coupling strengths, where higher temperatures (solid dark green) lead to a greater potential, as suggested by the intersection of the two curves. 
In Fig.~\ref{fig:work}(b), we show $\mathcal{W}$ as a function of temperature $k_{\mathrm{B}}T/\hbar\omega_{\mathrm{L}}$ for two coupling strengths $Q = 10.0\,\omega_{\mathrm{L}}\hbar^{-1}$ (solid dark green) and $Q = 25.0\,\omega_{\mathrm{L}}\hbar^{-1}$ (solid light green).
For low temperatures the limit of Eq.~\eqref{eq:work} is $\lim_{\beta \to \infty} \mathcal{W} = 0$. Likewise one has $\lim_{\beta \to 0} \mathcal{W} = 0$ because mean force corrections become less pronounced at higher temperatures~\cite{cerisola2022,cresser2021} and thus energy--shell inhomogeneities vanish in the ${\beta \to 0}$ limit.
This implies the existence of a maximum of the work extraction potential $\mathcal{W}$ at an intermediate temperature. 
I.e. there is a trade--off between the reorganization energy and the thermal energy. 

We conclude that some bath symmetries induce energy--shell inhomogeneities in the CMF state that can be linked to a work extraction potential $\mathcal{W}$. But we highlight that  \emph{not all} anisotropic baths generate such inhomogeneities. 

\section{Conclusion}
\label{sec:discussion}

In this paper, we showed that anisotropic three-dimensional baths acting on a classical spin vector $\mathbf{S}$ substantially modify its equilibrium state, the mean force state. Such baths arise whenever a spin is embedded in crystal lattices of varying symmetries, such as the orthorhombic or monoclinic symmetry. In addition to evaluating the mean force states directly, we numerically solved the system's steady state, demonstrating convergence to the mean force state.

Secondly, we identified the presence of a spin alignment transition in the classical spin--boson model. 
This is reminiscent of the well--known quantum phase transition, i.e. a change of the ground state at zero temperature, in the quantum spin--boson model.
In the classical case, we find that the bath symmetry determines whether this transition occurs smoothly (monoclinic) or abruptly (orthorhombic).

Thirdly, we demonstrated how inhomogeneous distributions of a classical open spin system, i.e. ``classical coherences'', lead to the presence of a thermodynamic work extraction potential, equivalent to their quantum counterpart. Here, the inhomogeneous nature of the CMF state is solely determined by the crystal symmetry. 
The upper limit of the work extraction potential depends on the coupling strength $Q$ and the bath temperature $T$. 

Understanding the impact of the symmetry of the surrounding environment is crucial to predict the equilibrium state of certain systems of interest. 
Examples of such systems include magnetic materials, such as thin cobalt films, where significantly different inertial spin dynamics have recently been observed for different crystal symmetries~\cite{unikandanunni2022}.
The presented results will also be relevant for the modelling of biological systems in highly complex environments~\cite{huelga2013}, such as photosynthesis~\cite{huelga2013,Trushechkin_2019}.

\section*{Acknowledgments}
FH thanks Franco Mayo, Augusto Roncaglia, Federico Cerisola, Charlie Hogg, and Joachim Ankerhold for helpful comments and discussions. SS is supported by a DTP grant from EPSRC (EP/R513210/1). FH and JA gratefully acknowledge funding from the Deutsche Forschungsgemeinschaft (DFG 513075417). JA gratefully acknowledges funding from EPSRC (EP/R045577/1) and thanks the Royal Society for support.

\bibliographystyle{unsrtnat}
\bibliography{main}

\begin{thebibliography}{58}
\providecommand{\natexlab}[1]{#1}
\providecommand{\url}[1]{\texttt{#1}}
\expandafter\ifx\csname urlstyle\endcsname\relax
  \providecommand{\doi}[1]{doi: #1}\else
  \providecommand{\doi}{doi: \begingroup \urlstyle{rm}\Url}\fi

\bibitem[Jarzynski(2004)]{jarzynski2004a}
Chris Jarzynski.
\newblock Nonequilibrium work theorem for a system strongly coupled to a
  thermal environment.
\newblock \emph{Journal of Statistical Mechanics: Theory and Experiment},
  2004\penalty0 (09):\penalty0 P09005, September 2004.
\newblock \doi{10.1088/1742-5468/2004/09/P09005}.

\bibitem[Seifert(2016)]{seifert2016}
Udo Seifert.
\newblock First and {{Second Law}} of {{Thermodynamics}} at {{Strong
  Coupling}}.
\newblock \emph{Physical Review Letters}, 116\penalty0 (2):\penalty0 020601,
  January 2016.
\newblock \doi{10.1103/PhysRevLett.116.020601}.

\bibitem[Jarzynski(2017)]{jarzynski2017a}
Christopher Jarzynski.
\newblock Stochastic and {{Macroscopic Thermodynamics}} of {{Strongly Coupled
  Systems}}.
\newblock \emph{Physical Review X}, 7\penalty0 (1):\penalty0 011008, January
  2017.
\newblock \doi{10.1103/PhysRevX.7.011008}.

\bibitem[Strasberg and Esposito(2017)]{strasberg2017}
Philipp Strasberg and Massimiliano Esposito.
\newblock Stochastic thermodynamics in the strong coupling regime: {{An}}
  unambiguous approach based on coarse graining.
\newblock \emph{Physical Review E}, 95\penalty0 (6):\penalty0 062101, June
  2017.
\newblock \doi{10.1103/PhysRevE.95.062101}.

\bibitem[Talkner and H{\"a}nggi(2020)]{talkner2020}
Peter Talkner and Peter H{\"a}nggi.
\newblock Colloquium: {{Statistical}} mechanics and thermodynamics at strong
  coupling: {{Quantum}} and classical.
\newblock \emph{Reviews of Modern Physics}, 92\penalty0 (4):\penalty0 041002,
  October 2020.
\newblock \doi{10.1103/RevModPhys.92.041002}.

\bibitem[Miller and Anders(2017)]{miller2017}
H.~J.~D. Miller and J.~Anders.
\newblock Entropy production and time asymmetry in the presence of strong
  interactions.
\newblock \emph{Physical Review E}, 95\penalty0 (6):\penalty0 062123, June
  2017.
\newblock \doi{10.1103/PhysRevE.95.062123}.

\bibitem[Aurell(2018)]{aurell2018}
Erik Aurell.
\newblock Unified picture of strong-coupling stochastic thermodynamics and time
  reversals.
\newblock \emph{Physical Review. E}, 97\penalty0 (4-1):\penalty0 042112, April
  2018.
\newblock \doi{10.1103/PhysRevE.97.042112}.

\bibitem[Trushechkin(2022)]{trushechkin2021}
Anton Trushechkin.
\newblock Quantum master equations and steady states for the
  ultrastrong-coupling limit and the strong-decoherence limit.
\newblock \emph{Phys. Rev. A}, 106:\penalty0 042209, Oct 2022.
\newblock \doi{10.1103/PhysRevA.106.042209}.

\bibitem[Thingna et~al.(2012)Thingna, Wang, and H{\"a}nggi]{thingna2012}
Juzar Thingna, Jian-Sheng Wang, and Peter H{\"a}nggi.
\newblock Generalized {{Gibbs}} state with modified {{Redfield}} solution:
  {{Exact}} agreement up to second order.
\newblock \emph{The Journal of Chemical Physics}, 136\penalty0 (19):\penalty0
  194110, May 2012.
\newblock \doi{10.1063/1.4718706}.

\bibitem[Campisi et~al.(2009)Campisi, Talkner, and H{\"a}nggi]{campisi2009}
Michele Campisi, Peter Talkner, and Peter H{\"a}nggi.
\newblock Thermodynamics and fluctuation theorems for a strongly coupled open
  quantum system: An exactly solvable case.
\newblock \emph{Journal of Physics A: Mathematical and Theoretical},
  42\penalty0 (39):\penalty0 392002, September 2009.
\newblock \doi{10.1088/1751-8113/42/39/392002}.

\bibitem[Trushechkin et~al.(2022)Trushechkin, Merkli, Cresser, and
  Anders]{trushechkin2022}
A.~S. Trushechkin, M.~Merkli, J.~D. Cresser, and J.~Anders.
\newblock Open quantum system dynamics and the mean force gibbs state.
\newblock \emph{AVS Quantum Science}, 4\penalty0 (1):\penalty0 012301, 2022.
\newblock \doi{10.1116/5.0073853}.

\bibitem[Binder et~al.(2018)Binder, Correa, Gogolin, Anders, and
  Adesso]{binder2018}
Felix Binder, Luis~A. Correa, Christian Gogolin, Janet Anders, and Gerardo
  Adesso, editors.
\newblock \emph{Thermodynamics in the {{Quantum Regime}}: {{Fundamental
  Aspects}} and {{New Directions}}}, volume 195 of \emph{Fundamental
  {{Theories}} of {{Physics}}}.
\newblock {Springer International Publishing}, {Cham}, 2018.
\newblock \doi{10.1007/978-3-319-99046-0}.

\bibitem[Chiu et~al.(2022)Chiu, Strathearn, and Keeling]{chiu2022}
Yiu-Fung Chiu, Aidan Strathearn, and Jonathan Keeling.
\newblock Numerical evaluation and robustness of the quantum mean-force
  {{Gibbs}} state.
\newblock \emph{Physical Review A}, 106\penalty0 (1):\penalty0 012204, July
  2022.
\newblock \doi{10.1103/PhysRevA.106.012204}.

\bibitem[Cresser and Anders(2021)]{cresser2021}
J.~D. Cresser and J.~Anders.
\newblock Weak and {{Ultrastrong Coupling Limits}} of the {{Quantum Mean Force
  Gibbs State}}.
\newblock \emph{Physical Review Letters}, 127\penalty0 (25):\penalty0 250601,
  December 2021.
\newblock \doi{10.1103/PhysRevLett.127.250601}.

\bibitem[{Anto-Sztrikacs} et~al.(2022){Anto-Sztrikacs}, Nazir, and
  Segal]{anto-sztrikacs2022a}
Nicholas {Anto-Sztrikacs}, Ahsan Nazir, and Dvira Segal.
\newblock Effective {{Hamiltonian}} theory of open quantum systems at strong
  coupling.
\newblock \emph{arXiv:2211.05701}, November 2022.
\newblock \doi{10.48550/arXiv.2211.05701}.

\bibitem[Cerisola et~al.(2022)Cerisola, Berritta, Scali, Horsley, Cresser, and
  Anders]{cerisola2022}
Federico Cerisola, Marco Berritta, Stefano Scali, Simon A.~R. Horsley, James~D.
  Cresser, and Janet Anders.
\newblock Quantum-classical correspondence in spin-boson equilibrium states at
  arbitrary coupling.
\newblock \emph{arXiv:2204.10874}, April 2022.
\newblock \doi{10.48550/arXiv.2204.10874}.

\bibitem[Smith et~al.(2022)Smith, Sinha, and Jarzynski]{smith2022}
Andrew Smith, Kanupriya Sinha, and Christopher Jarzynski.
\newblock Quantum {{Coherences}} and {{Classical Inhomogeneities}} as
  {{Equivalent Thermodynamics Resources}}.
\newblock \emph{Entropy}, 24\penalty0 (4):\penalty0 474, April 2022.
\newblock \doi{10.3390/e24040474}.

\bibitem[P{\k{e}}ka{\l}a et~al.(2013)P{\k{e}}ka{\l}a, {Wolff-Fabris}, Fagnard,
  Vanderbemden, Mucha, Gospodinov, Lovchinov, and Ausloos]{pekala2013}
M.~P{\k{e}}ka{\l}a, F.~{Wolff-Fabris}, J-F. Fagnard, {\relax Ph}~Vanderbemden,
  J.~Mucha, M.~M. Gospodinov, V.~Lovchinov, and M.~Ausloos.
\newblock Magnetic properties and anisotropy of orthorhombic {{DyMnO3}} single
  crystal.
\newblock \emph{Journal of Magnetism and Magnetic Materials}, 335:\penalty0
  46--52, June 2013.
\newblock \doi{10.1016/j.jmmm.2013.01.036}.

\bibitem[Harikrishnan et~al.(2009)Harikrishnan, R{\"o}{\ss}ler, Kumar, Bhat,
  R{\"o}{\ss}ler, Wirth, Steglich, and Elizabeth]{harikrishnan2009}
S.~Harikrishnan, S.~R{\"o}{\ss}ler, C.~M.~Naveen Kumar, H.~L. Bhat, U.~K.
  R{\"o}{\ss}ler, S.~Wirth, F.~Steglich, and Suja Elizabeth.
\newblock Phase transitions and rare-earth magnetism in hexagonal and
  orthorhombic {{DyMnO3}} single crystals.
\newblock \emph{Journal of Physics: Condensed Matter}, 21\penalty0
  (9):\penalty0 096002, February 2009.
\newblock \doi{10.1088/0953-8984/21/9/096002}.

\bibitem[Bhoi et~al.(2019)Bhoi, Dam, Mohapatra, Patidar, Singh, Singh,
  Vishwakarma, Babu, Siruguri, and Pradhan]{bhoi2019}
Krishnamayee Bhoi, Tapabrata Dam, S.~R. Mohapatra, Manju~Mishra Patidar,
  Durgesh Singh, A.~K. Singh, P.~N. Vishwakarma, P.~D. Babu, V.~Siruguri, and
  {\relax Dillip. K}~Pradhan.
\newblock Studies of magnetic phase transitions in orthorhombic {{DyMnO3}}
  ceramics prepared by acrylamide polymer gel template method.
\newblock \emph{Journal of Magnetism and Magnetic Materials}, 480:\penalty0
  138--149, June 2019.
\newblock \doi{10.1016/j.jmmm.2019.02.069}.

\bibitem[Bogdanov et~al.(2007)Bogdanov, Zhuravlev, and
  R{\"o}{\ss}ler]{bogdanov2007a}
A.~N. Bogdanov, A.~V. Zhuravlev, and U.~K. R{\"o}{\ss}ler.
\newblock Spin-flop transition in uniaxial antiferromagnets: {{Magnetic}}
  phases, reorientation effects, and multidomain states.
\newblock \emph{Physical Review B}, 75\penalty0 (9):\penalty0 094425, March
  2007.
\newblock \doi{10.1103/PhysRevB.75.094425}.

\bibitem[Antropov et~al.(2021)Antropov, Kravtsov, Makarova, Proglyado, Keller,
  Subbotin, Pashaev, Prutskov, Vasiliev, Chesnokov, Bebenin, Milyaev, Ustinov,
  Keimer, and Khaydukov]{antropov2021}
N.~O. Antropov, E.~A. Kravtsov, M.~V. Makarova, V.~V. Proglyado, T.~Keller,
  I.~A. Subbotin, E.~M. Pashaev, G.~V. Prutskov, A.~L. Vasiliev, {\relax Yu.
  M}~Chesnokov, N.~G. Bebenin, M.~A. Milyaev, V.~V. Ustinov, B.~Keimer, and
  {\relax Yu. N}~Khaydukov.
\newblock Tunable spin-flop transition in artificial ferrimagnets.
\newblock \emph{Physical Review B}, 104\penalty0 (5):\penalty0 054414, August
  2021.
\newblock \doi{10.1103/PhysRevB.104.054414}.

\bibitem[Biswas et~al.(2022)Biswas, Michel, Fjellv{\aa}g, Bimashofer,
  D{\"o}beli, Jambor, Keller, M{\"u}ller, Ukleev, Pomjakushina, Singh, Stuhr,
  Vaz, Lippert, and Schneider]{biswas2022}
Banani Biswas, Veronica~F. Michel, {\O}ystein~S. Fjellv{\aa}g, Gesara
  Bimashofer, Max D{\"o}beli, Michal Jambor, Lukas Keller, Elisabeth
  M{\"u}ller, Victor Ukleev, Ekaterina~V. Pomjakushina, Deepak Singh, Uwe
  Stuhr, C.~A.~F. Vaz, Thomas Lippert, and Christof~W. Schneider.
\newblock Role of {{Dy}} on the magnetic properties of orthorhombic dyfeo3.
\newblock \emph{Physical Review Materials}, 6\penalty0 (7):\penalty0 074401,
  July 2022.
\newblock \doi{10.1103/PhysRevMaterials.6.074401}.

\bibitem[Srivastava(2020)]{srivastava2020}
S.~K. Srivastava.
\newblock Magnetic {{Property}} of {{Mn-Doped Monoclinic ZrO2 Compounds}}.
\newblock \emph{Journal of Superconductivity and Novel Magnetism}, 33\penalty0
  (8):\penalty0 2501--2505, August 2020.
\newblock \doi{10.1007/s10948-020-05522-1}.

\bibitem[Long et~al.(2011)Long, Zhang, Li, Sabirianov, Zhang, and
  Zeng]{long2011}
Gen Long, Hongwang Zhang, Da~Li, Renat Sabirianov, Zhidong Zhang, and Hao Zeng.
\newblock Magnetic anisotropy and coercivity of {{Fe3Se4}} nanostructures.
\newblock \emph{Applied Physics Letters}, 99\penalty0 (20):\penalty0 202103,
  November 2011.
\newblock \doi{10.1063/1.3662388}.

\bibitem[Zhang et~al.(2011)Zhang, Long, Li, Sabirianov, and Zeng]{zhang2011}
Hongwang Zhang, Gen Long, Da~Li, Renat Sabirianov, and Hao Zeng.
\newblock {{Fe3Se4 Nanostructures}} with {{Giant Coercivity Synthesized}} by
  {{Solution Chemistry}}.
\newblock \emph{Chemistry of Materials}, 23\penalty0 (16):\penalty0 3769--3774,
  August 2011.
\newblock \doi{10.1021/cm201610k}.

\bibitem[Singh et~al.(2020)Singh, Gupta, He, and Sonvane]{singh2020}
Deobrat Singh, Sanjeev~K. Gupta, Haiying He, and Yogesh Sonvane.
\newblock First-principles study of the electronic, magnetic and optical
  properties of {{Fe3Se4}} in its monoclinic phase.
\newblock \emph{Journal of Magnetism and Magnetic Materials}, 498:\penalty0
  166157, March 2020.
\newblock \doi{10.1016/j.jmmm.2019.166157}.

\bibitem[Anders et~al.(2007)Anders, Bulla, and Vojta]{anders2007a}
Frithjof~B. Anders, Ralf Bulla, and Matthias Vojta.
\newblock Equilibrium and {{Nonequilibrium Dynamics}} of the {{Sub-Ohmic
  Spin-Boson Model}}.
\newblock \emph{Physical Review Letters}, 98\penalty0 (21):\penalty0 210402,
  May 2007.
\newblock \doi{10.1103/PhysRevLett.98.210402}.

\bibitem[Alvermann and Fehske(2009)]{alvermann2009}
A.~Alvermann and H.~Fehske.
\newblock Sparse polynomial space approach to dissipative quantum systems:
  Application to the sub-ohmic spin-boson model.
\newblock \emph{Phys. Rev. Lett.}, 102:\penalty0 150601, Apr 2009.
\newblock \doi{10.1103/PhysRevLett.102.150601}.

\bibitem[Florens et~al.(2010)Florens, Venturelli, and Narayanan]{florens2010}
S.~Florens, D.~Venturelli, and R.~Narayanan.
\newblock Quantum phase transition in the spin boson model.
\newblock In \emph{Quantum Quenching, Annealing and Computation}, pages
  145--162. Springer Berlin Heidelberg, 2010.
\newblock \doi{10.1007/978-3-642-11470-0\_6}.

\bibitem[De~Filippis et~al.(2020{\natexlab{a}})De~Filippis, {de Candia},
  Cangemi, Sassetti, Fazio, and Cataudella]{defilippis2020a}
G.~De~Filippis, A.~{de Candia}, L.~M. Cangemi, M.~Sassetti, R.~Fazio, and
  V.~Cataudella.
\newblock Quantum {{Phase Transitions}} in the {{Spin-Boson}} model:
  {{MonteCarlo Method}} vs {{Variational Approach}} a la {{Feynman}}.
\newblock \emph{Physical Review B}, 101\penalty0 (18):\penalty0 180408, May
  2020{\natexlab{a}}.
\newblock \doi{10.1103/PhysRevB.101.180408}.

\bibitem[Wang et~al.(2019)Wang, He, Duan, and Chen]{wang2019b}
Yan-Zhi Wang, Shu He, Liwei Duan, and Qing-Hu Chen.
\newblock Quantum phase transitions in the spin-boson model without the
  counterrotating terms.
\newblock \emph{Physical Review B}, 100\penalty0 (11):\penalty0 115106,
  September 2019.
\newblock \doi{10.1103/PhysRevB.100.115106}.

\bibitem[Leggett et~al.(1987)Leggett, Chakravarty, Dorsey, Fisher, Garg, and
  Zwerger]{leggett1987}
A.~J. Leggett, S.~Chakravarty, A.~T. Dorsey, Matthew P.~A. Fisher, Anupam Garg,
  and W.~Zwerger.
\newblock Dynamics of the dissipative two-state system.
\newblock \emph{Reviews of Modern Physics}, 59\penalty0 (1):\penalty0 1--85,
  January 1987.
\newblock \doi{10.1103/RevModPhys.59.1}.

\bibitem[Weiss(1999)]{weiss1999}
Ulrich Weiss.
\newblock \emph{Quantum {{Dissipative Systems}}}.
\newblock {World Scientific}, 1999.
\newblock \doi{10.1142/6738}.

\bibitem[Ferialdi(2017)]{ferialdi2017}
L.~Ferialdi.
\newblock Exact non-{{Markovian}} master equation for the spin-boson and
  {{Jaynes-Cummings}} models.
\newblock \emph{Physical Review A}, 95\penalty0 (2):\penalty0 020101, February
  2017.
\newblock \doi{10.1103/PhysRevA.95.020101}.

\bibitem[Orman and Kawai(2020)]{orman2020}
Patrick~Lee Orman and Ryoichi Kawai.
\newblock A qubit strongly interacting with a bosonic environment: {{Geometry}}
  of thermal states.
\newblock \emph{arXiv}, October 2020.
\newblock \doi{10.48550/arXiv.2010.09201}.

\bibitem[Lloyd(2011)]{lloyd2011}
Seth Lloyd.
\newblock Quantum coherence in biological systems.
\newblock \emph{Journal of Physics: Conference Series}, 302:\penalty0 012037,
  July 2011.
\newblock \doi{10.1088/1742-6596/302/1/012037}.

\bibitem[Kolli et~al.(2012)Kolli, O'Reilly, Scholes, and
  {Olaya-Castro}]{kolli2012}
Avinash Kolli, Edward~J. O'Reilly, Gregory~D. Scholes, and Alexandra
  {Olaya-Castro}.
\newblock The fundamental role of quantized vibrations in coherent light
  harvesting by cryptophyte algae.
\newblock \emph{The Journal of Chemical Physics}, 137\penalty0 (17):\penalty0
  174109, November 2012.
\newblock \doi{10.1063/1.4764100}.

\bibitem[Huelga and Plenio(2013)]{huelga2013}
S.F. Huelga and M.B. Plenio.
\newblock Vibrations, quanta and biology.
\newblock \emph{Contemporary Physics}, 54\penalty0 (4):\penalty0 181--207, July
  2013.
\newblock \doi{10.1080/00405000.2013.829687}.

\bibitem[Arndt et~al.(2009)Arndt, Juffmann, and Vedral]{arndt2009}
Markus Arndt, Thomas Juffmann, and Vlatko Vedral.
\newblock Quantum physics meets biology.
\newblock \emph{HFSP Journal}, 3\penalty0 (6):\penalty0 386--400, December
  2009.
\newblock \doi{10.2976/1.3244985}.

\bibitem[Dattagupta(2021)]{dattagupta2021}
Sushanta Dattagupta.
\newblock Spin-boson model of quantum dissipation in graphene: {{Nonlinear}}
  electrical response.
\newblock \emph{Physical Review B}, 104\penalty0 (8):\penalty0 085411, August
  2021.
\newblock \doi{10.1103/PhysRevB.104.085411}.

\bibitem[Wang et~al.(2017)Wang, Ren, and Cao]{wang2017a}
Chen Wang, Jie Ren, and Jianshu Cao.
\newblock Unifying quantum heat transfer in a nonequilibrium spin-boson model
  with full counting statistics.
\newblock \emph{Physical Review A}, 95\penalty0 (2):\penalty0 023610, February
  2017.
\newblock \doi{10.1103/PhysRevA.95.023610}.

\bibitem[De~Filippis et~al.(2020{\natexlab{b}})De~Filippis, {de Candia},
  Cangemi, Sassetti, Fazio, and Cataudella]{defilippis2020}
G.~De~Filippis, A.~{de Candia}, L.~M. Cangemi, M.~Sassetti, R.~Fazio, and
  V.~Cataudella.
\newblock Quantum phase transitions in the spin-boson model: {{Monte Carlo}}
  method versus variational approach a la {{Feynman}}.
\newblock \emph{Physical Review B}, 101\penalty0 (18):\penalty0 180408, May
  2020{\natexlab{b}}.
\newblock \doi{10.1103/PhysRevB.101.180408}.

\bibitem[Zhang et~al.(2010)Zhang, Chen, and Wang]{zhang2010}
Yu-Yu Zhang, Qing-Hu Chen, and Ke-Lin Wang.
\newblock Quantum phase transition in the sub-{{Ohmic}} spin-boson model:
  {{An}} extended coherent-state approach.
\newblock \emph{Physical Review B}, 81\penalty0 (12):\penalty0 121105, March
  2010.
\newblock \doi{10.1103/PhysRevB.81.121105}.

\bibitem[Dzsotjan et~al.(2010)Dzsotjan, S{\o}rensen, and
  Fleischhauer]{dzsotjan2010}
David Dzsotjan, Anders~S. S{\o}rensen, and Michael Fleischhauer.
\newblock Quantum emitters coupled to surface plasmons of a nanowire: {{A
  Green}}'s function approach.
\newblock \emph{Physical Review B}, 82\penalty0 (7):\penalty0 075427, August
  2010.
\newblock \doi{10.1103/PhysRevB.82.075427}.

\bibitem[Nemati et~al.(2022)Nemati, Henkel, and Anders]{nemati2022}
S.~Nemati, C.~Henkel, and J.~Anders.
\newblock Coupling function from bath density of states.
\newblock \emph{Europhysics Letters}, 139\penalty0 (3):\penalty0 36002, July
  2022.
\newblock \doi{10.1209/0295-5075/ac7b42}.

\bibitem[Anders et~al.(2022)Anders, Sait, and Horsley]{anders2022}
J.~Anders, C.~R.~J. Sait, and S.~A.~R. Horsley.
\newblock Quantum {{Brownian}} motion for magnets.
\newblock \emph{New Journal of Physics}, 24\penalty0 (3):\penalty0 033020,
  March 2022.
\newblock \doi{10.1088/1367-2630/ac4ef2}.

\bibitem[Mori and Miyashita(2008)]{mori2008}
Takashi Mori and Seiji Miyashita.
\newblock Dynamics of the {{Density Matrix}} in {{Contact}} with a {{Thermal
  Bath}} and the {{Quantum Master Equation}}.
\newblock \emph{Journal of the Physical Society of Japan}, 77\penalty0
  (12):\penalty0 124005, December 2008.
\newblock \doi{10.1143/JPSJ.77.124005}.

\bibitem[Scali et~al.(2023)Scali, Horsley, Anders, and Cerisola]{scali}
Stefano Scali, Simon Horsley, Janet Anders, and Federico Cerisola.
\newblock Spidy.jl -- open--source {J}ulia package for the study of anisotropic
  system--bath interaction. \emph{In Preparation.}, 2023.

\bibitem[Neumann and Meyer(1885)]{neumann1885}
Franz~Ernst Neumann and Oskar~Emil Meyer.
\newblock \emph{{Vorlesungen \"uber die Theorie der Elastizit\"at der festen
  K\"orper und des Licht\"athers}}.
\newblock {B.G. Teubner}, 1885.

\bibitem[Hauss{\"u}h(2007)]{haussuh2007}
Siegfried Hauss{\"u}h.
\newblock \emph{Physical {{Properties}} of {{Crystals}} - {{An}} Introduction}.
\newblock {WILEY - VCH Verlag GmbH\&Co. KGaA}, {Weinheim}, 2007.
\newblock ISBN 978-3-527-40543-5.

\bibitem[Malgrange et~al.(2014)Malgrange, Ricolleau, and
  Schlenker]{malgrange2014}
C{\'e}cile Malgrange, Christian Ricolleau, and Michel Schlenker.
\newblock \emph{Symmetry and {{Physical Properties}} of {{Crystals}}}.
\newblock {Springer Netherlands}, {Dordrecht}, 2014.
\newblock \doi{10.1007/978-94-017-8993-6}.

\bibitem[Newnham(2005)]{newnham2005}
Robert~E. Newnham.
\newblock \emph{Properties of Materials: Anisotropy, Symmetry, Structure}.
\newblock {Oxford University Press}, {Oxford ; New York}, 2005.
\newblock \doi{10.1093/oso/9780198520757.001.0001}.

\bibitem[Zhou et~al.(2015)Zhou, Chen, Xu, Chernyak, and Zhao]{Zhou_2015}
Nengji Zhou, Lipeng Chen, Dazhi Xu, Vladimir Chernyak, and Yang Zhao.
\newblock Symmetry and the critical phase of the two-bath spin-boson model:
  Ground-state properties.
\newblock \emph{Physical Review B}, 91\penalty0 (19), May 2015.
\newblock \doi{10.1103/physrevb.91.195129}.

\bibitem[Xu et~al.(2022)Xu, Yan, Shi, Ankerhold, and Stockburger]{xu2022}
Meng Xu, Yaming Yan, Qiang Shi, J.~Ankerhold, and J.~T. Stockburger.
\newblock Taming {{Quantum Noise}} for {{Efficient Low Temperature
  Simulations}} of {{Open Quantum Systems}}.
\newblock \emph{Physical Review Letters}, 129\penalty0 (23):\penalty0 230601,
  November 2022.
\newblock \doi{10.1103/PhysRevLett.129.230601}.

\bibitem[Kammerlander and Anders(2016)]{kammerlander2016}
P.~Kammerlander and J.~Anders.
\newblock Coherence and measurement in quantum thermodynamics.
\newblock \emph{Scientific Reports}, 6\penalty0 (1):\penalty0 22174, February
  2016.
\newblock \doi{10.1038/srep22174}.

\bibitem[Unikandanunni et~al.(2022)Unikandanunni, Medapalli, Asa, Albisetti,
  Petti, Bertacco, Fullerton, and Bonetti]{unikandanunni2022}
Vivek Unikandanunni, Rajasekhar Medapalli, Marco Asa, Edoardo Albisetti,
  Daniela Petti, Riccardo Bertacco, Eric~E. Fullerton, and Stefano Bonetti.
\newblock Inertial {{Spin Dynamics}} in {{Epitaxial Cobalt Films}}.
\newblock \emph{Physical Review Letters}, 129\penalty0 (23):\penalty0 237201,
  November 2022.
\newblock \doi{10.1103/PhysRevLett.129.237201}.

\bibitem[Trushechkin(2019)]{Trushechkin_2019}
Anton Trushechkin.
\newblock Calculation of coherences in förster and modified redfield theories
  of excitation energy transfer.
\newblock \emph{The Journal of Chemical Physics}, 151\penalty0 (7):\penalty0
  074101, August 2019.
\newblock \doi{10.1063/1.5100967}.

\end{thebibliography}

\appendix
\section{Derivation of the classical mean force state}
\label{app:MFder}
\subsection{Classical spin trace}
The coordinates of a spin in a spherical coordinate system are given as $(S_x,S_y,S_z) = (S_0\sin\vartheta\cos\varphi,S_0\sin\vartheta\sin\varphi,S_0\cos\vartheta)$ with $\vartheta \in [0,\pi]$, $\varphi \in [0,2\pi)$, and a vector length of $S_0$.
The trace of a function $A(S_x,S_y,S_z)$ is
\begin{align}
\mathrm{tr_{S}^{cl}}[A(S_x,S_y,S_z)] &= \nonumber\\
\frac{1}{4\pi}\int_0^{2\pi}\mathrm{d}\varphi&\int_0^{\pi}\mathrm{d}\vartheta\,\sin\vartheta A(S_x,S_y,S_z).
\end{align}
The definition of $\tilde{Z}_{\mathrm{S}}^{\mathrm{cl}}$ follows trivially.

\subsection{Classical bath trace}
\label{subsec:bathtrace}
The trace over the bath degrees of freedom is calculated via a discrete version of the bath Hamiltonian $H_{\mathrm{B}} = \frac{1}{2}\sum_n \left(\mathbf{P}_{\omega_n}^2+\omega_n^2\mathbf{X}^2_{\omega_n}\right)$.
The bath partition function becomes,
\begin{align}
    Z_{\mathrm{B}}^{\mathrm{cl}} &= \prod_{n}\left[\int_{-\infty}^{\infty}\mathrm{d}\mathbf{X}_{\omega_n}\int_{-\infty}^{\infty}\mathrm{d}\mathbf{P}_{\omega_n}
    e^{-\frac{1}{2}\beta\left(\mathbf{P}^2_{\omega_n}+\omega_n^2\mathbf{X}_{\omega_n}^2 \right)}\right]
    \label{eq:bathtrace}
\end{align}

\subsection{Mean force state}
Following Ref.~\cite{cerisola2022}, the mean force in the three--dimensional setting is calculated via the  discretise bath degrees of freedom (see~\ref{subsec:bathtrace}), such that $H_{\mathrm{tot}}$ becomes
\begin{align}
    H_{\mathrm{tot}} = H_{\mathrm{S}} + \sum_{n=0}^{\infty}\left[\frac{1}{2}\Big(\mathbf{P}_{\omega_{n}}^{2} + \omega_n^2\mathbf{X}^2_{\omega_n}\Big) - \mathbf{S}C_{\omega_n}\mathbf{X}_{\omega_n}\right].
\end{align}
We can simplify the integration over the bath degrees of freedom (see Eq.~\eqref{eq:bathtrace}), by completing the square, via $\mu_n = \mathbf{S}\mathcal{C}_{\omega_n}$,
\begin{align}
    H_{\mathrm{tot}} = H_{\mathrm{S}}+\sum_{n = 0}^{\infty}\left[\frac{1}{2}\left[\mathbf{P}^2_{\omega_n}+\omega^2_n\Big(\mathbf{X}_{\omega_n}-\frac{\mathbf{\mu}_{\omega_n}}{\omega_n^2}\Big)^2\right]-\frac{\mathbf{\mu}_{\omega_n}^2}{2\omega_n^2}\right]
\end{align}
The classical system--bath partition function is,
\begin{equation}
Z_{\mathrm{SB}}^{\mathrm{cl}} = \int_{0}^{2\pi}\mathrm{d}\varphi\int_{0}^{\pi}\mathrm{d}\vartheta\sin\vartheta e^{-\beta H_{\mathrm{eff}}}Z_{\mathrm{B}}^{\mathrm{cl}},
\end{equation}
whereby we define the effective Hamiltonian as,
\begin{align}
    H_{\mathrm{eff}} = H_{\mathrm{S}} - Q\tilde{S}^2.
\end{align}
The reorganization energy $Q$ and $\tilde{S}$ are defined in the main text (see~\ref{sec:setup}).
This yields the partition function of the system,
\begin{align}
\tilde{Z}_{\mathrm{S}}^{\mathrm{cl}} = \frac{Z_{\mathrm{SB}}^{\mathrm{cl}}}{Z_{\mathrm{B}}^{\mathrm{cl}}} = \int_{0}^{2\pi}\mathrm{d}\varphi\int_0^{\pi}\mathrm{d}\vartheta \sin\vartheta e^{-\beta H_{\mathrm{eff}}},
\end{align}
and additionally the mean force state of the three--dimensional spin--boson system:
\begin{align}
\tau_{\mathrm{MF}} = \frac{1}{\tilde{Z}_{\mathrm{S}}^{\mathrm{cl}}} e^{-\beta H_{\mathrm{eff}}}.
\end{align}

\subsection{Classical Expectation Values}
The expectation value is defined by $\langle S_i \rangle = \int \mathrm{d}\Omega\,S_i\,\tau_{\mathrm{MF}}$, with $\mathrm{d}\Omega = \sin\vartheta\,\mathrm{d}\vartheta\,\mathrm{d}\varphi$ and $i \in (x,y,z)$.
We normalize the expectation value with respect to the spin length, i.e. $\langle s_i \rangle = \langle S_i \rangle/S_0$, such that:
\begin{align}
\langle s_x \rangle &= \frac{1}{\tilde{Z}_{\mathrm{S}}^{\mathrm{cl}}} \int_0^{2\pi}\mathrm{d}\varphi\cos\varphi \int_0^\pi\mathrm{d}\vartheta \sin^2\vartheta\,e^{\left[-\beta H_{\mathrm{eff}}(\vartheta,\varphi)\right]} \\
\langle s_y \rangle&= \frac{1}{\tilde{Z}_{\mathrm{S}}^{\mathrm{cl}}} \int_0^{2\pi}\mathrm{d}\varphi\sin\varphi \int_0^\pi\mathrm{d}\vartheta \sin^2\vartheta\,e^{\left[-\beta H_{\mathrm{eff}}(\vartheta,\varphi)\right]} \\
\langle s_z \rangle &= \frac{1}{\tilde{Z}_{\mathrm{S}}^{\mathrm{cl}}} \int_0^{2\pi}\mathrm{d}\varphi \int_0^\pi\mathrm{d}\vartheta \sin\vartheta\cos\vartheta\,e^{\left[-\beta H_{\mathrm{eff}}(\vartheta,\varphi)\right]}.
\end{align}

\section{Bumpy expectation value vs. temperature plot}
\label{subsec:bumpyexp}
\begin{figure}
    \centering
    \includegraphics[width=0.47\textwidth]{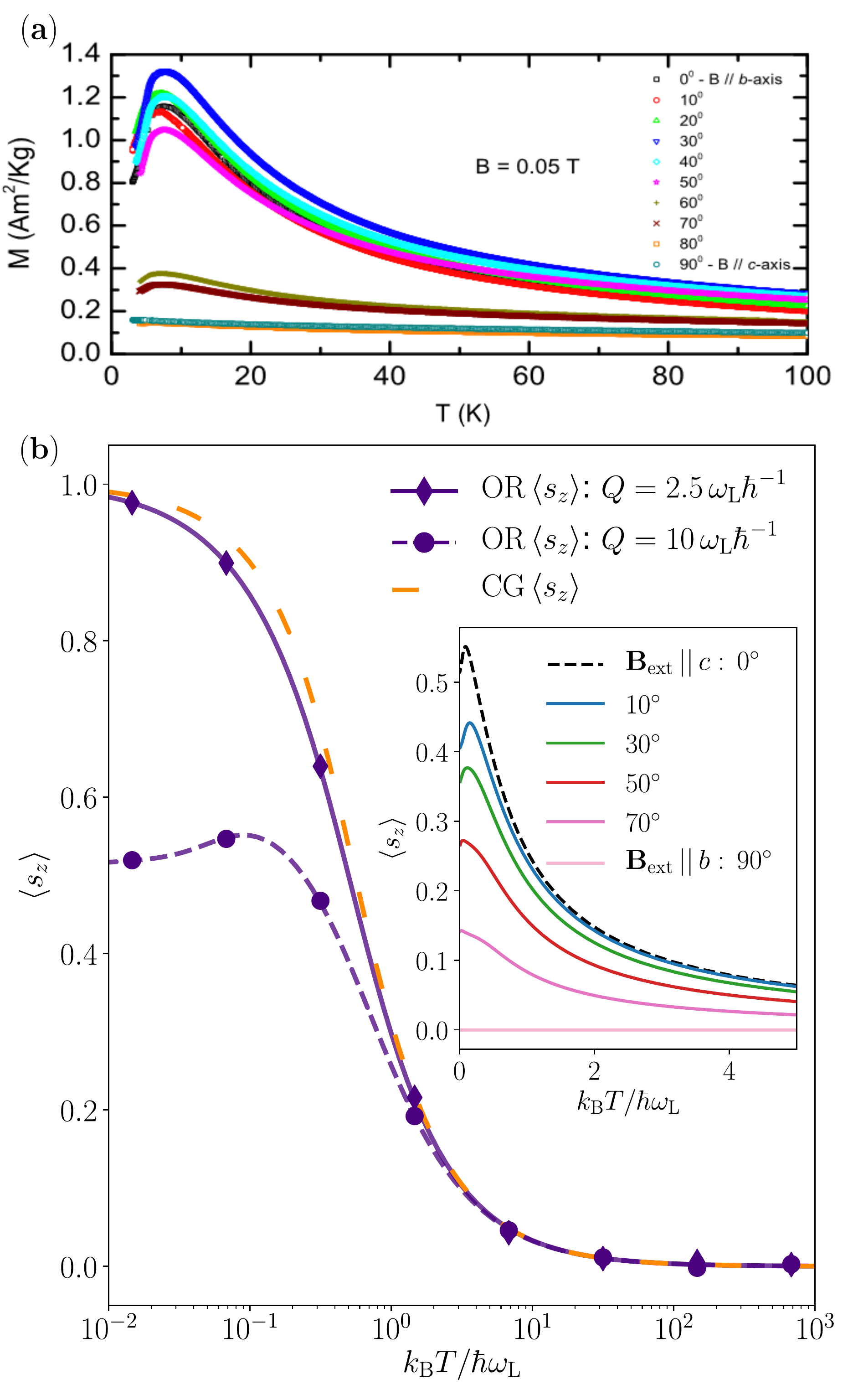}
    \caption{\textbf{Bumpy expectation value:} (a)
    DC magnetization measurements were conducted by P{\k{e}}ka{\l}a et al.~\cite{pekala2013} for a single crystal orthorhombic DyMnO3. The figure shows the 
  temperature dependence of the magnetization for different magnetic field orientations. (Figure taken from \textit{JMMM} \textbf{335}, 46-52 (2013).)
 (b) Our calculated expectation value of the orthorhombic (OR) CMF state (Eq.~\eqref{eq:meanforce0}) at two different coupling strengths $Q = 2.5\,\omega_{\mathrm{L}}\hbar^{-1}$ and $Q = 10.0\,\omega_{\mathrm{L}}\hbar^{-1}$. At strong reorganization energies the expectation value reaches a maximum far above zero temperature (e.g. at $Q = 10.0\,\omega_{\mathrm{L}}\hbar^{-1}$ the maximal magnetization is at $k_{\mathrm{B}}T/\hbar\omega_{\mathrm{L}} \approx 0.1$).
 The inset shows the expectation value for different magnetic field orientations, similar to the experiment shown in (a).}
    \label{fig:OR_bump}
\end{figure}

As mentioned in the main text, at large system--bath coupling the magnetization is maximized above the $T \to 0\,\mathrm{K}$ limit.
We show the bumpy expectation value in Fig.~\ref{fig:OR_bump}(b).
In addition, we simulate the CSS via Refs.~\cite{anders2022,scali}, which reproduces the bumpy expectation value (see Fig.~\ref{fig:OR_bump}(b)). 
The bump directly results from the anisotropy of the bath.
In fact, there is a competition between the energy term of the system $H_{\mathrm{S}}$ and the mean force correction term $Q\tilde{S}^2$. 
The external magnetic field tries to align the spin in its direction, whereas the correction term has other preferred directions. 
In the low temperature range the environment overcomes the system energy and reduces the expectation value $\langle s_z \rangle$ to a value below its maximum.
We encounter this bumpy behaviour solely for the orthorhombic symmetry and for a coupling tensor~\eqref{eq:proptensor}, where $c_{12}\neq0$ and $c_{11}$, $c_{22}$, $c_{33}$ are as in the orthorhombic case.

As we have pointed out in the main text, similar magnetization curves are observed in experiments~\cite{pekala2013,harikrishnan2009,bhoi2019,bogdanov2007a,biswas2022}.
Here we show the temperature dependent magnetization measurements from Ref.~\cite{pekala2013} (see Fig.~\ref{fig:OR_bump}(a)) for an orthorhombic DyMnO3 single crystal for different external magnetic field orientations. 
The inset in Fig.~\ref{fig:OR_bump}(b) shows a similar behaviour of the expectation value of the orthorhombic CMF state for different external magnetic field orientations.
We leave a more detailed study for future work.

\section{Inhomogeneities}
\label{sec:appendix_inhomogeneities}
The orthorhombic CMF state is, 
\begin{align}
    \tau_{\mathrm{MF}}^{\mathrm{OR}}&(\vartheta,\varphi) = \label{eq:orthoMF}\\
    \frac{1}{\tilde{Z}_{\mathrm{S}}^{\mathrm{cl}}}&e^{-\beta \left(H_{\mathrm{S}} - QS_0^2((c_{11}\cos^2\varphi + c_{22}\sin^2\varphi)\sin^2\vartheta + c_{33}\cos^2\vartheta)\right)},\nonumber
\end{align}
with $c_{11} \neq c_{22} \neq c_{33}$. 
Following Ref.~\cite{smith2022}, we have to calculate $\eta(E) = \int\mathrm{d}\Omega\,\tau_{\mathrm{MF}}\,\delta(E-H_{\mathrm{S}}(\vartheta))$.
It is helpful to consider the following integral identity,
\begin{equation}
    \int_0^{2\pi} \mathrm{d}\varphi\,\exp\big(a\cos^2\varphi\big) = 2\pi e^{a/2}\mathrm{I}_0\bigg(\frac{a}{2}\bigg),
\end{equation}
where $\mathrm{I}_0$ is the modified Bessel function of first kind.
Using $\cos^2\varphi + \sin^2\varphi = 1$, we can rewrite Eq.~\eqref{eq:orthoMF} and identify the parameter $a = \beta QS_0^2\sin^2\vartheta(c_{11}-c_{22})$, such that
\begin{align}
    \eta(E) = \frac{2\pi}{\tilde{Z}_{\mathrm{S}}^{\mathrm{cl}}}&\int_0^{\pi} \mathrm{d}\vartheta\sin\vartheta\,\mathrm{I}_0\left(\frac{a}{2}\right)\delta\big(E - H_{\mathrm{S}}(\vartheta)) \label{eq:AppB1}\\
    &e^{\left[\beta\omega_{\mathrm{L}}S_0\cos\vartheta + \frac{1}{2}\beta Q S_0^2 \sin^2\vartheta(c_{11}+c_{22}) + \beta Q c_{33}\cos^2\vartheta\right]}. \nonumber
\end{align}
The $\vartheta$--integration is straightforward due to $\delta\big(E - H_{\mathrm{S}}(\vartheta))$. With the substitutions $u = \omega_{\mathrm{L}}S_0\cos\vartheta$, $\mathrm{d}u = -\omega_{\mathrm{L}}S_0\sin\vartheta\mathrm{d}\vartheta$, and the properties of the delta function, Eq.~\eqref{eq:AppB1} is,
\begin{align}
    \eta(E) =& \frac{2\pi}{\tilde{Z}_{\mathrm{S}}^{\mathrm{cl}}}\frac{1}{\omega_{\mathrm{L}}S_0}e^{\left[-\beta E + \beta Q\omega_{\mathrm{L}}^{-2}\left(\frac{1}{2}(1-E^2)(c_{11}+c_{22}) + c_{33}E^2\right)\right]} \nonumber \\
    &\times\Theta(\omega_{\mathrm{L}}S_0+E)\Theta(\omega_{\mathrm{L}}S_0-E)\\
    &\times\mathrm{I}_0\left(\frac{1}{2}\beta Q \omega_{\mathrm{L}}^{-2}(c_{11} - c_{22})(1 - E^2)\right)\nonumber,
\end{align}
where $\Theta(\cdot)$ is the step--function. 
The classical microcanonical density $\omega_E(\vartheta)$ is~\cite{smith2022},
\begin{equation}
    \omega_E(\vartheta) = \frac{\omega_{\mathrm{L}}S_0}{2\pi}\frac{\delta\left(E - H_{\mathrm{S}}(\vartheta)\right)}{\Theta(\omega_{\mathrm{L}}S_0+E)\Theta(\omega_{\mathrm{L}}S_0-E)}
\end{equation}
The homogeneous phase space distribution $\mathrm{diag}\,\tau_{\mathrm{MF}}^{\mathrm{OR}}(\vartheta)$ is defined as $\mathrm{diag}\,\tau_{\mathrm{MF}}^{\mathrm{OR}}(\vartheta) = \int\mathrm{d}E\,\eta(E)\omega_{\mathrm{E}}(\vartheta,\varphi)$ (see Ref.~\cite{smith2022}).
This leads to, 
\begin{align}
    \mathrm{diag}\,\tau_{\mathrm{MF}}^{\mathrm{OR}}(\vartheta) &= \frac{1}{\tilde{Z}_{\mathrm{S}}^{\mathrm{cl}}}\mathrm{I}_0\left(\frac{a}{2}\right) \\
    &e^{\left[-\beta\left(H_{\mathrm{S}} - \frac{1}{2}QS_0^2\sin^2\vartheta(c_{11} + c_{22}) - QS_0^2c_{33}\cos^2\vartheta \right)\right]} \nonumber.
\end{align}
We observe that $\mathrm{diag}\,\tau_{\mathrm{MF}}^{\mathrm{OR}}(\vartheta,\varphi)\neq \tau_{\mathrm{MF}}^{\mathrm{OR}}(\vartheta)$, therefore by Eq.~\eqref{eq:work}, $\mathcal{W} \geq 0$.

\end{document}